\documentclass[11pt]{article}
\usepackage{amsmath,amssymb,bbm,amsthm}
\usepackage{fullpage}
\usepackage{thm-restate,color,xcolor,xspace}
\usepackage{hyperref,cleveref}
\usepackage{graphicx}
\usepackage{algorithm,algorithmic}

\newtheorem{theorem}{Theorem}
\newtheorem{lemma}[theorem]{Lemma}

\begin{document}

\title{Deterministic Padded Decompositions \\ and Negative-Weight Shortest Paths}
\author{Jason Li\footnote{Carnegie Mellon University. email: \tt jmli@cs.cmu.edu}}
\date{\today}
\maketitle

\begin{abstract}
We obtain the first near-linear time deterministic algorithm for negative-weight single-source shortest paths on integer-weighted graphs. Our main ingredient is a deterministic construction of a padded decomposition on directed graphs, which may be of independent interest.
\end{abstract}

\section{Introduction}

We consider the problem of computing single-source shortest paths on a directed, edge-weighted graph. While a near-optimal algorithm on graphs with non-negative weights has been known since Dijkstra, the setting with negative weights has remained open since Bellman-Ford in the 1950s, and only recently has the problem been settled on graphs with negative \emph{integral} weights~\cite{bernstein2025negative}. This breakthrough of Bernstein, Nanongkai, and Wulff-Nilsen combined the integral \emph{scaling} approach~\cite{goldberg95} with directed low-diameter graph decompositions~\cite{leighton1999multicommodity}, establishing the first near-linear time algorithm for the problem with integral weights. Since then, the running time has been improved to $O((m+n\log\log n)\log(nW)\log n\log\log n)$~\cite{bringmann2023negative,li2026faster}, but all of these algorithms are \emph{randomized}, relying crucially on (probabilistic) low-diameter decompositions. It remained an open problem whether a \emph{deterministic} algorithm exists in near-linear time, which we answer affirmatively in this work.

\begin{theorem}\label{thm:main}
Given a graph with integral edge weights at least $-W$, there is a deterministic algorithm that computes single-source shortest paths in $O((m+n\log\log n)\log(nW)\log^3n)$ time.
\end{theorem}

We remark that a recent trend of considering \emph{real-valued} weights has also witnessed exciting breakthroughs \cite{fineman2024single,huang2025faster,huang2026faster}, beating the $O(mn)$ algorithm of Bellman-Ford for the first time in 70~years. Finally, another recent line of work on using \emph{continuous optimization} for graph algorithms has also featured multiple breakthroughs~\cite{chen2025maximum,van2023deterministic}, including a deterministic $m^{1+o(1)}\log W$ time algorithm for the more general problem of minimum-cost flow.

\paragraph{Independent work.}
In an independent work, Haeupler, Jiang, and Saranurak~\cite{haeupler2025deterministic} also obtain a near-linear time deterministic algorithm for negative-weight shortest paths. Their algorithm defines and constructs a \emph{path cover}, a derandomization of directed low-diameter decompositions that may have future applications. In comparison, our algorithm is faster and more direct.

\subsection{Our Techniques}

Our main insight is to replace the low-diameter decompositions from prior work, which are inherently probabilistic, with a \emph{padded} graph decomposition technique that may be of independent interest. At a high level, given a radius parameter $\Delta$, the algorithm seeks to find a vertex set $U$ such that the \emph{padding} $P$ consisting of all vertices in $V\setminus U$ of distance at most $\Delta/\log^2n$ from $U$ has a relatively small size. Moreover, the partition $(U,V\setminus U)$ should be balanced, in that both sides support a constant fraction of the graph size. (In general, it may not be possible to find such a set $U$, but assume for now that we can.) The algorithm makes two recursive calls to negative-weight single-source shortest paths, one on $G[U\cup P]$ and one on $G[V\setminus U]$. Observe that vertices in the padding participate in both recursive calls, which is why we require the padding to be small. Importantly, the padding forces any path that visits $U$ to spend another $\Delta/\log^2n$ distance within $U\cup P$, which means it cannot enter $U$ and then escape $U\cup P$ an arbitrary number of times. This idea is inspired by the concept of \emph{fringe} vertices from Fineman's parallel shortcutting algorithm~\cite{fineman2018nearly}, where vertices are similarly duplicated in recursive calls. After the two recursive calls on $G[U\cup P]$ and $G[V\setminus U]$, the algorithm constructs an auxiliary graph where all shortest paths \emph{deterministically} contain few negative-weight edges, in comparison to prior work which only guarantees it in expectation. This step assumes the scaling framework of~\cite{bernstein2025negative} and may output a negative-weight cycle instead. One interesting difference in our algorithm is that we completely eliminate the need to fix DAG edges, since our padded decomposition does not treat DAGs as a special base case.

Of course, finding such a set $U$ with small padding may be impossible. For example, if the graph itself has diameter at most $\Delta/\log^2n$, then the padding for any vertex set $U$ is simply $V\setminus U$, and since we enforce the set $V\setminus U$ to support a constant fraction of the graph, the padding cannot be sub-linear in size. In general, the algorithm is also allowed to find a large subgraph of small \emph{weak} diameter, similar to prior randomized algorithms~\cite{fischer2025simple,li2026faster}. Also, for technical reasons, we may define the padding in either direction, i.e., either vertices of distance at most $\Delta/\log^2n$ out of $U$, or vertices of distance at most $\Delta/\log^2n$ into $U$, and we also obtain up to three sets in the decomposition, not two.

\subsection{Preliminaries}

Let $G=(V,E,w)$ be a directed graph with integral edge weights. The degree of a vertex $v\in V$ is the number of edges incident to $v$ (which does not depend on edge directions or weights), and the volume $\textup{vol}(U)$ of a vertex subset $U\subseteq V$ equals $\sum_{v\in U}\deg(v)$. For vertices $u,v\in V$, define $d(u,v)$ to be the distance from $u$ to $v$ according to the edge weights. %More generally, for a vertex subset $U\s V$, define $d(U,v)=\min_{u\in U}d(u,v)$.

The following definitions of weak diameter and in/out-ball are exclusively used on non-negative weighted graphs.
For a vertex subset $U\subseteq V$, define the \emph{weak diameter} of $U$ in $G$ as $\max\{d(u,v):u,v\in U\}$. For vertex $s\in V$ and non-negative number $r$, define the out-ball $B^+_G(s,r)=\{v\in V:d(s,v)\le r\}$ and in-ball $B^-_G(s,r)=\{v\in V:d(v,s)\le r\}$. More generally, for a vertex subset $U\subseteq V$ and a non-negative number $r$, define the out-ball $B^+_G(U,r)=\{v\in V:\min_{s\in U}d(s,v)\le r\}$ and in-ball $B^-_G(U,r)=\{v\in V:\min_{s\in U}d(v,s)\le r\}$.

A \emph{potential function} $\phi:V\to\mathbb R$ is a function used to re-weight the edges of the graph, where each edge $(u,v)$ has new weight $w(u,v)+\phi(u)-\phi(v)$. A potential is \emph{valid} if the new edge weights are all non-negative. By a result of Johnson~\cite{johnson1977efficient}, the task of computing negative-weight single-source shortest paths reduces to finding a valid potential function, after which a single call to Dijkstra is made.

The iterative \emph{scaling} technique of~\cite{bernstein2025negative} does not compute such a potential function in one go. Instead, if the graph $G$ has all edge weights at least $-W$, then the goal is to either compute a potential function such that the new edge weights are at least $-W/2$, or output a negative-weight cycle. After $O(\log(nW))$ iterations, all edge weights are at least $-1/n^{O(1)}$, at which point the single-source shortest paths can be recovered by rounding.

In order to compute this potential function, the algorithm of~\cite{bernstein2025negative} first defines the graph $G^{W/2}$ as the graph $G$ with value $W/2$ added to all edge weights. Then, any valid potential function for $G^{W/2}$ is a potential function for $G$ where all new edge weights are at least $-W/2$. For the rest of this paper, our task is to compute a valid potential function for $G^{W/2}$ in $O((m+n\log\log n)\log^3n)$ deterministic time, which suffices for \Cref{thm:main}.

Similar to prior work, we run our decomposition algorithm on the graph $G^{W/2}_{\ge0}$ defined as the graph $G^{W/2}$ with all negative-weight edges replaced by zero-weight edges. Finally, we need a Bellman-Ford/Dijkstra hybrid algorithm~\cite{dinitz2017hybrid}, also used in prior work. In fact, our Bellman-Ford/Dijkstra is even simpler because all of our relevant shortest paths have few negative-weight edges \emph{deterministically}, not just in expectation. %For convenience, we do not mention a source vertex in this shortest paths subroutine. Instead, the goal is to compute, for each vertex $v\in V$, the shortest path ending at $v$ (and starting from anywhere). Note that the shortest path has weight at most $0$ since the empty path at $v$ of weight $0$ is a possible choice.
\begin{lemma}[Hybrid Bellman-Ford/Dijkstra]\label{lem:BFD}
For any parameter $\eta\ge1$, there is a deterministic $O(\eta\cdot(m+n\log\log n))$ time algorithm that outputs one of the following:
 \begin{enumerate}
 \item The correct shortest path distances from the source $s$ to each vertex $v\in V$, or
% \im A negative-weight cycle, or
 \item A path $P$ from $s$ to some vertex $v\in V$ that is shorter than any $s$-to-$v$ path with at most $\eta$ negative-weight edges.
 \end{enumerate}
\end{lemma}

\section{Negative-Weight Shortest Paths}

We begin with the statement of our padded decomposition algorithm on graphs with non-negative weights, which replaces the probabilistic low-diameter decomposition from prior work. For a set $U$ with padding $P$ described in the introduction, imagine that $k=2$ and the two sets $X_1,X_2$ are $U\cup P$ and $V\setminus U$, so they share $P$ in common. The concept of padding itself is replaced by the path coloring statement in property~(\ref{item:3}). For a set $U$ with padding $P$, imagine coloring the vertices on a path sequentially as follows: the first vertex on the path inside $U$ is colored~$1$, and all subsequent vertices are also colored~$1$ until the path leaves $U\cup P$, at which point the vertices are colored~$2$ until the next entrance to $U$. The statement guarantees that the path has few edges with differently colored endpoints, which coincides with the intuition that $P$ cannot enter $U$ and leave $U\cup P$ too many times.

\begin{restatable}[Padded decomposition]{theorem}{Padded}\label{thm:padded-decomposition}
Let $m$ be larger than some constant, and let $d,\epsilon>0$ be parameters, where $\epsilon$ is smaller than some constant $\epsilon_0>0$. Consider a graph $G$ on $n$ vertices with $m$ non-negative weight edges. There is a deterministic $O(m+n\log\log n)$ time algorithm that finds sets $X_1,\ldots,X_k$ with $k\le3$ such that
\begin{enumerate}
\item For each set $X_i$, either $\textup{vol}(X_i)\le1.6m$ or $X_i$ has weak diameter at most $d/2$.\label{item:1}
\item $X_1\cup\cdots\cup X_k=V$ and $\textup{vol}(X_1)+\cdots+\textup{vol}(X_k)\le 2m+O(\epsilon m)$.\label{item:2}
\item Consider any path of weight at most $d$. The vertices of this path can be assigned colors in $\{1,\ldots,k\}$ such that
 \begin{enumerate}
 \item For all $i\in\{1,\ldots,k\}$, each vertex colored $i$ is in $X_i$.
 \item At most $100\epsilon^{-1}\log m$ edges on the path have differently colored endpoints.
 \end{enumerate}\label{item:3}
\end{enumerate}
\end{restatable}

We defer our padded decomposition algorithm to \Cref{sec:padded-decomposition}. For the rest of this section, we use it to solve negative-weight single-source shortest paths. We prove the theorem below, which combined with the scaling framework of~\cite{bernstein2025negative} establishes \Cref{thm:main}.
\begin{theorem}\label{thm:single-iteration-of-scaling}
Given a graph $G$ with integral edge weights at least $-W$, there is a deterministic $O((m+n\log\log n)\log^3n)$ time algorithm that either computes single-source shortest paths on $G^{W/2}$, or outputs a negative-weight cycle in $G$.
\end{theorem}

The algorithm runs the following recursive procedure starting with $X=V$ and $d=m^2W/2$. The output is a valid potential function on $G^{W/2}[X]$, which is enough to solve single-source shortest paths on $G^{W/2}$ when $X=V$. The recursive algorithm keeps the parameter $\epsilon$ fixed through the recursive calls, which we eventually set as $\epsilon=1/\log m$.

\paragraph{The algorithm on graph $G^{W/2}$ with recursive parameters $(X,d)$:}
\begin{enumerate}
\item (Base case.) If the number of edges in $G[X]$ is less than some constant, then output a potential function in constant time.
\item (Base case.) If $d<W/2$:
 \begin{enumerate}
 \item If there is a negative-weight edge in $G^{W/2}[X]$, then find and output a negative-weight cycle in $G$.\label{item:2a2}
 \item Otherwise, output the potential function that is $0$ everywhere.
 \end{enumerate}
\item Run \Cref{thm:padded-decomposition} on the graph $G^{W/2}_{\ge0}[X]$ with $m'$ edges and parameter $\epsilon=1/\log m$, obtaining sets $X_1,\ldots,X_k$. Here, we assume that $m$ is large enough to satisfy $\epsilon=1/\log m<\epsilon_0$ from \Cref{thm:padded-decomposition}.
\item Recursively solve $(X_1,d_1),\ldots,(X_k,d_k)$, where $d_i=d$ if $\textup{vol}(X_i)\le1.6m'$ and $d_i=d/2$ otherwise, and let $\phi_1,\ldots,\phi_k$ be the resulting potential functions.
\item Construct the following auxiliary graph $H$:
 \begin{enumerate}
 \item The vertex set is the \emph{disjoint} union $\{s\}\uplus X_1\uplus\cdots\uplus X_k$. For index $i\in\{1,\ldots,k\}$ and vertex $v\in X_i$, let $v_i$ be the copy of vertex $v$ from set $X_i$.
 \item For each vertex $v\in X$ and index $i$ such that $v\in X_i$, add the edge $(s,v_i)$ with weight $-\phi_i(v)$.
 \item For each edge $(u,v)$ and indices $i,j$ such that $u\in X_i$ and $v\in X_j$, add the edge $(u_i,v_j)$ with weight $w(u,v)+\phi_i(u)-\phi_j(v)$.
 \end{enumerate}
\item Run Bellman-Ford/Dijkstra (\Cref{lem:BFD}) on $H$ with source $s$ and $\eta=100\log^2m+1$. If shortest path distances are not returned, then find and output a negative-weight cycle in $G$.
\item Output the following potential function $\phi$ on $G^{W/2}[X]$: for each $v\in X$, set $\phi(v)=d_H(s,v_i)+\phi_i(v)$ for any index $i$ with $v\in X_i$.
\end{enumerate}

For the rest of this section, we prove correctness and establish the running time. All lemmas below assume that $(X,d)$ is an instance of the recursive algorithm.

\begin{lemma}\label{lem:weak-diameter}
If $d<m^2W/2$, then $X$ has weak diameter at most $d$ in $G^{W/2}_{\ge0}$.
\end{lemma}
\begin{proof}
Consider going down the recursion tree from the root to instance $(X,d)$, and let $(X',d')$ be the first instance satisfying $d'=d$. Since $d\ne m^2W/2$, it cannot be the root instance, so its own parent instance made the recursive call $(X',d)$ with $\textup{vol}(X')>1.6m''$, where $m''$ is the number of edges of this parent instance. By property~(\ref{item:1}) of \Cref{thm:padded-decomposition}, $X'$ has weak diameter at most $d$ in $G^{W/2}_{\ge0}$. Since $X\subseteq X'$, we conclude that $X$ also has weak diameter at most $d$ in $G^{W/2}_{\ge0}$.
\end{proof}
\begin{lemma}
If step~(\ref{item:2a2}) is executed, then the algorithm can find a negative-weight cycle in $G$.
\end{lemma}
\begin{proof}
Consider a negative-weight edge $(u,v)$ in $G^{W/2}[X]$, which has weight less than $-W/2$ in $G$. By \Cref{lem:weak-diameter}, there is a path from $v$ to $u$ in $G^{W/2}$ of weight at most $d<W/2$, which can be found by Dijkstra's algorithm. This path also has weight less than $W/2$ in $G$, and adding the edge $(u,v)$ forms a negative-weight cycle in $G$.
\end{proof}
\begin{lemma}\label{lem:path-mapping}
Any $s$-to-$v_i$ path $Q$ in $H$ corresponds to a path $P$ in $G^{W/2}[X]$ ending at $v$ (and starting from anywhere), and  the weight of $Q$ is lower by exactly $\phi_i(v)$. In particular, the final potential function $\phi$ is valid in $G^{W/2}[X]$.
\end{lemma}
\begin{proof}
%Imagine first weighting the edges $(u_i,v_j)$ of $H$ by their original weight $w(u,v)$, and edges incident to $s$ have weight $0$. By property~(\ref{item:2}) of \thm{padded-decomposition}, we have $X_1\cup\cds\cup X_k=X$, so all paths in $G^{W/2}[X]$ are captured by $H$.

Consider an $s$-to-$v_i$ path $Q$ with vertices $s,v^{(1)}_{i_1},\ldots,v^{(k)}_{i_k}=v_i$. The weight of $Q$ in $H$ is
\[ -\phi_{i_1}(v^{(1)})+\big((w_G(v^{(1)},v^{(2)})+\phi_{i_1}(v^{(1)})-\phi_{i_2}(v^{(2)})\big)+\cdots+\big(w_G(v^{(k-1)},v^{(k)})+\phi_{i_{k-1}}(v^{(k-1)})-\phi_{i_k}(v^{(k)})\big) ,\]
which telescopes to $w_G(v^{(1)},v^{(2)})+\cdots+w_G(v^{(k-1)},v^{(k)})-\phi_i(v)$. So the path $Q$ corresponds to the path $P$ in $G^{W/2}[X]$ with vertices $v^{(1)},\ldots,v^{(k)}=v$, and the weight of $Q$ is offset by exactly $-\phi_i(v)$. In particular, since $\phi(v)=d_H(s,v_i)+\phi_i(v)$ exactly undoes this offset for the shortest $s$-to-$v_i$ path, it captures original distances in $G^{W/2}$ (starting from anywhere), so it is valid.
\end{proof}
\begin{lemma}\label{lem:path-mapping2}
Any path $P$ in $G^{W/2}[X]$ ending at $v$ with weight at most $d$ in $G^{W/2}_{\ge0}$ corresponds to an $s$-to-$v_i$ path $Q$ in $H$ with at most $100\log^2m+1$ negative-weight edges in $G^{W/2}$, and the weight of $Q$ is lower by exactly $\phi_i(v)$.
\end{lemma}
\begin{proof}
Consider a shortest path $P$ ending at $v$ of weight at most $d$ in $G^{W/2}_{\ge0}$. By \Cref{thm:padded-decomposition}, there is a coloring of the vertices of $P$ with at most $100\epsilon^{-1}\log m'\le100\log^2m$ differently colored edges. Consider mapping $P$ to the path $Q$ in $H$ where a vertex $v$ in $P$ colored $i$ is mapped to vertex $v_i$ in $H$, and the vertex $s$ is added in front. The path $Q$ has at most $100\log^2m$ potentially negative edges $(u_i,v_j)$ with $i\ne j$, and one potentially negative edge out of $s$. The other edges $(u_i,v_i)$ are non-negative because each $G^{W/2}[X_i]$ has non-negative edges under the valid potential function $\phi_i$. It follows that the path $Q$ has at most $100\log^2m+1$ negative-weight edges.

Finally, let $v_i$ be the last vertex on the path $Q$. Similarly to the proof of \Cref{lem:path-mapping}, the weight of $Q$ is lower by exactly $\phi_i(v)$.
\end{proof}

\begin{lemma}
If Bellman-Ford/Dijkstra returns a path $Q$ to a vertex in $H$, then the algorithm can find a negative-weight cycle in $G$.
\end{lemma}
\begin{proof}
If Bellman-Ford/Dijkstra returns an $s$-to-$v_i$ path $Q$ in $H$, then it is shorter than any $s$-to-$v_i$ path with at most $\eta$ negative-weight edges. The path consisting of the single edge $(s,v_i)$ is such a path, so $Q$ has weight less than $w_H(s,v_i)=-\phi_i(v)$. By \Cref{lem:path-mapping}, there is a corresponding path $P$ in $G^{W/2}$ of weight less than $0$. If $P$ has weight at most $d$ in $G^{W/2}_{\ge0}$, then \Cref{lem:path-mapping2} guarantees a path $Q'$ in $H$ of the same weight as $Q$ with at most $\eta=100\log^2m+1$ negative-weight edges, contradicting the guarantee of $Q$. It follows that $P$ has weight more than $d$ in $G^{W/2}_{\ge0}$. Moreover, since $P$ has negative weight overall in $G^{W/2}$, its total negative weight (i.e., sum of negative-weight edges) in $G^{W/2}$ is less than the negative of its total positive weight, which is $-w_{G^{W/2}_{\ge0}}(P)<-d$.

First, consider the initial case $d=m^2W/2$. Since Bellman-Ford/Dijkstra runs in at most $m^2$ time when $m$ is large enough, the path $P$ itself has at most $m^2$ edges. Each negative-weight edge in $G^{W/2}$ has weight at least $-W/2$, and $P$ contains at most $m^2$ negative-weight edges, so its total negative weight in $G^{W/2}$ is at least $-W/2\cdot m^2=-d$, a contradiction.

For the rest of the proof, assume that $d<m^2W/2$. Recall that $P$ has total negative weight less than $-w_{G^{W/2}_{\ge0}}(P)$ in $G^{W/2}$. For each edge $(u,v)$ on $P$ with $w_{G^{W/2}}(u,v)<0$, its weight in $G$ is $w_{G^{W/2}}(u,v)-W/2\le2w_{G^{W/2}}(u,v)$, so the total negative weight of $P$ in $G$ is less than double the total negative weight of $P$ in $G^{W/2}$. In total, $P$ has total negative weight less than $-2w_{G^{W/2}_{\ge0}}(P)$ in $G$. Since $G$ has smaller edge weights than $G^{W/2}$, the path $P$ also has total positive weight at most $w_{G^{W/2}_{\ge0}}(P)$ in $G$, so the overall weight of $P$ is less than $-w_{G^{W/2}_{\ge0}}(P)<-d$ in $G$. Since $X$ has weak diameter at most $d$ in $G^{W/2}_{\ge0}$, we can close the path $P$ into a cycle with an additional weight of at most $d$ in $G^{W/2}_{\ge0}$ (as well as in $G$), forming a negative-weight cycle. This additional path of weight at most $d$ in $G^{W/2}_{\ge0}$ can be computed by Dijkstra's algorithm.
\end{proof}

\begin{lemma}
The total running time is $O((m+n\log\log n)\log^3n)$.
\end{lemma}
\begin{proof}
By property~(\ref{item:1}) of~\Cref{thm:padded-decomposition}, in each recursive call, the product $d\cdot\textup{vol}(X)$ decreases by a constant factor. Since $d$ ranges from $m^2W/2$ down to $W/2$, the recursion depth is $O(\log m)$. By property~(\ref{item:2}), the total size of recursive instances increases by factor $O(\epsilon)=O(1/\log m)$ per level of recursion, so the total number of edges is $m\cdot(1+O(1/\log m))^{O(\log m)}=O(m)$ per level. By regularizing vertex degrees, e.g., by adding $m/n$ self-loops of weight $0$ at each vertex, we can ensure that there are also $O(n)$ total vertices per level. On a recursive instance with $n'$ vertices and $m'$ edges, \Cref{thm:padded-decomposition} runs in $O(m'+n'\log\log n)$ time, and Bellman-Ford/Dijkstra runs in $O((m'+n'\log\log n)\log^2n)$ time. Summing over all $O(\log n)$ levels, the total running time is $O((m+n\log\log n)\log^3n)$.
\end{proof}

\section{Padded Decomposition}\label{sec:padded-decomposition}

In this section, we present our padded decomposition algorithm to prove \Cref{thm:padded-decomposition}, restated below.

\Padded*

We begin with a simple procedure to find in-balls and out-balls with small padding, which uses standard ball-growing arguments.

\begin{lemma}\label{lem:find-ball}
Given a graph $G=(V,E)$, a source vertex $s\in V$, a direction $\pm\in\{+,-\}$, an initial radius parameter $\Delta_0\ge0$, an additional radius limit $\Delta>0$, and a parameter $\epsilon>0$, there is an algorithm that computes a radius $r\in[\Delta_0,\Delta_0+\Delta]$ such that $\textup{vol}(B^\pm(s,r+\epsilon\Delta/\log m))\le(1+O(\epsilon))\,\textup{vol}(B^\pm(s,r))$, and runs in $O(\textup{vol}(B^\pm(s,r+\epsilon\Delta/\log m))+|B^\pm(s,r+\epsilon\Delta/\log m)|\log\log n)$ time.
\end{lemma}
\begin{proof}
Without loss of generality, assume that $\pm$ is $+$. The algorithm grows the ball $B^+(s,r)$ by running Dijkstra's algorithm, maintaining the current radius $r$ and ball $B^+(s,r)$. Each time $r$ surpasses $\Delta_0+i\cdot\epsilon\Delta/\log m$ for a positive integer $i$, the algorithm checks whether
\[ \textup{vol}(B^+(s,\Delta_0+i\cdot\epsilon\Delta/\log m))\le(1+C\epsilon)\,\textup{vol}(B^+(s,\Delta_0+(i-1)\cdot\epsilon\Delta/\log m)) \]
for a fixed constant $C>0$, and if so, immediately terminate with the radius $r=\Delta_0+(i-1)\cdot\epsilon\Delta/\log m$. The algorithm can check this condition in $O(1)$ time by maintaining the following history: for each index $j\le\textup{vol}(B^+(s,r))$ in an array, store the minimum radius $r'$ for which $\textup{vol}(B^+(s,r'))\ge j$. The algorithm simply looks up index $\lceil\textup{vol}(B^+(s,\Delta_0+i\cdot\epsilon\Delta/\log m))/(1+C\epsilon)\rceil$ and compares it to $\Delta_0+(i-1)\cdot\epsilon\Delta/\log m$.

For the running time, we implement Dijkstra's algorithm using Thorup's priority queue~\cite{thorup2003integer}. Each vertex update takes $O(\log\log n)$ time and each edge update takes $O(1)$ time, and the running time follows.

Finally, we prove that $r\le\Delta_0+\Delta$ by a standard ball-growing argument. If the algorithm does not stop by positive integer $i$, then
\begin{align*}
\textup{vol}(B^+(s,\Delta_0+i\cdot\epsilon\Delta/\log m))&>(1+C\epsilon)\,\textup{vol}(B^+(s,\Delta_0+(i-1)\cdot\epsilon\Delta/\log m))
\\&>(1+C\epsilon)^2\,\textup{vol}(B^+(s,\Delta_0+(i-2)\cdot\epsilon\Delta/\log m))
\\&>\cdots
\\&>(1+C\epsilon)^i\,\textup{vol}(B^+(s,\Delta_0))
\\&\ge(1+C\epsilon)^i.
\end{align*}
Since $\textup{vol}(B^+(s,\Delta_0+i\cdot\epsilon\Delta/\log m))\le 2m$, we conclude that $2m\ge(1+C\epsilon)^i$. For constant $C>0$ large enough, the algorithm must stop by $i\le\epsilon^{-1}\log m$, so $r=\Delta_0+(i-1)\cdot\epsilon\Delta/\log m\le\Delta_0+\Delta$.
\end{proof}

\subsection{Light case}

Suppose that \Cref{lem:find-ball} always finds either an in-ball or out-ball with at most half the volume. In this case, we stop the algorithm once the total volume of in-balls or out-balls covers a constant fraction of the volume, and set $U$ to be the union of these balls. If the algorithm fails to find a small in-ball or out-ball from a vertex $s$, then the algorithm gives up on the light case, proceeding to the heavy case in \Cref{sec:heavy-case} with the vertex $s$.

In the algorithm below, imagine that $\Delta\approx d$; we will eventually set $\Delta=d/12$ to handle all cases. Below, we define $G\{U\}$ as the induced subgraph $G[U]$ with (zero-weight) self-loops added to preserve vertex degrees, i.e., $\deg_{G\{U\}}(v)=\deg_G(v)$ for all $v\in U$.
\begin{enumerate}
\item Initialize $U^+\gets\emptyset$ and $U^-\gets\emptyset$.
\item While $\textup{vol}(U^+)<m/2$ and $\textup{vol}(U^-)<m/2$:
 \begin{enumerate}
 \item Select an arbitrary vertex $s\in V\setminus(U^+\cup U^-)$. Run \Cref{lem:find-ball} on $G\{V\setminus U^+\}$ in the $+$ direction, and separately on $G\{V\setminus U^-\}$ in the $-$ direction, both with initial radius $0$ and additional limit $\Delta$. Run the two algorithms in parallel, and let $B^\pm_{G\{V\setminus U^\pm\}}(s,r)$ be the ball of smaller volume, where $\pm\in\{+,-\}$. Terminate the other algorithm immediately upon finding the smaller ball.\label{item:2a}
 \item If $\textup{vol}(B^\pm_{G\{V\setminus U^\pm\}}(s,r))\le m$, then update $U^\pm\gets U^\pm\cup B^\pm_{G\{V\setminus U^\pm\}}(s,r)$.
 \item Else, terminate the light case immediately and proceed to the heavy case with the vertex $s$.
 \end{enumerate}
\item Return either $U^+$ or $U^-$, whichever has larger volume.
\end{enumerate}

In the proofs below, we assume that the light case does not terminate early (with the heavy case). Define $X=B^\pm_G(U^\pm,\epsilon\Delta/\log m)$ and $Y=V\setminus U^\pm$.

\begin{lemma}\label{lem:size-light}
We have $\textup{vol}(X)\le1.6m$ and $\textup{vol}(Y)\le3m/2$ and $\textup{vol}(X)+\textup{vol}(Y)\le 2m+O(\epsilon m)$.
\end{lemma}
\begin{proof}
Without loss of generality, assume that $\pm$ is $+$. We have $\textup{vol}(U^+)\ge m/2$ since the while loop terminated, so $\textup{vol}(Y)\le3m/2$. On the last iteration of the while loop, we had $\textup{vol}(U^+)<m/2$ before adding in a set of volume at most $m$, so $\textup{vol}(U^+)\le3m/2$ at the end, which means $\textup{vol}(Y)\ge m/2$. Suppose first that we have proved $\textup{vol}(X)+\textup{vol}(Y)\le 2m+O(\epsilon m)$; then $\textup{vol}(X)\le 2m-\textup{vol}(Y)+O(\epsilon m)\le 3m/2+O(\epsilon m)$, which is at most $1.6m$ since $\epsilon<\epsilon_0$ for a small enough constant $\epsilon_0>0$.

To prove $\textup{vol}(X)+\textup{vol}(Y)\le 2m+O(\epsilon m)$, it suffices to prove that their intersection\linebreak $B^+_G(U^+,\epsilon\Delta/\log m)\setminus U^+$ has volume at most $O(\epsilon m)$. Define the \emph{padded} vertices as the union of the sets $B^+_{G\{V\setminus U^+\}}(s,r+\epsilon\Delta/\log m)\setminus B^+_{G\{V\setminus U^+\}}(s,r)$ over all balls $B^+_{G\{V\setminus U^+\}}(s,r)$ that were added to $U^+$. By the guarantee of \Cref{lem:find-ball},
\[ \textup{vol}(B^+_{G\{V\setminus U^+\}}(s,r+\epsilon\Delta/\log m))\le(1+O(\epsilon))\,\textup{vol}(B^+_{G\{V\setminus U^+\}}(s,r)), \]
so the volume ratio of newly padded vertices to new vertices in $U^+$ is $O(\epsilon)$. It follows that the total volume of padded vertices is $O(\epsilon m)$. To complete the proof, we show that each vertex $v\in B^+_G(U^+,\epsilon\Delta/\log m)\setminus U^+$ is a padded vertex. Consider such a vertex $v$, and consider the shortest path $P$ from $U^+$ to $v$ of weight at most $\epsilon\Delta/\log m$. Let $u$ be the vertex on $P$ that was added to $U^+$ the earliest over the course of the algorithm. Then, on the iteration immediately before adding $u$ to $U^+$, the $u$-to-$v$ segment of $P$ is contained in $G[V\setminus U^+]$ and also has weight at most $\epsilon\Delta/\log m$. Since $u\in B^+_{G\{V\setminus U^+\}}(s,r)$ on this iteration, we have $v\in B^+_{G\{V\setminus U^+\}}(s,r+\epsilon\Delta/\log m)$. In addition, $v\notin B^+_{G\{V\setminus U^+\}}(s,r)$ since $v$ was never added to $U^+$, so $v$ is a padded vertex.
\end{proof}

\begin{lemma}\label{lem:coloring-light}
Consider any path of weight $w$. The vertices of this path can be colored red and blue such that
 \begin{enumerate}
 \item Each red vertex is in $X$, and each blue vertex is in $Y$.
 \item At most $2\lceil\frac w{\epsilon\Delta/\log m}\rceil$ edges on the path have differently colored endpoints.
 \end{enumerate}
\end{lemma}
\begin{proof}
Suppose first that $\pm$ is $+$, and let the path be $P$. Consider the first vertex on $P$ that is in $U^+$, and call it $u$. Consider the maximal segment of vertices after $u$ that are all in $X=B^\pm_G(U^\pm,\epsilon\Delta/\log m)$, and let $v$ be the last vertex on this segment. If there is a vertex $v'$ after $v$ on $P$, then the $u$-to-$v'$ segment of $P$ must have weight exceeding $\epsilon\Delta/\log m$.

We color everything before $u$ blue, and everything from $u$ to $v$ red. If the vertex $v'$ exists, then we recursively color everything from $v'$ onwards. Observe that on each recursive call, the weight of $P$ decreases by more than $\epsilon\Delta/\log m$, so there are at most $\lceil\frac w{\epsilon\Delta/\log m}\rceil$ total calls. Each iteration has at most $2$ edges with different colors, which concludes the proof.

The case when $\pm$ is $-$ is identical, except we start at the end of the path and color backwards.
\end{proof}

\begin{lemma}
The algorithm runs in $O(m+n\log\log n)$ time.
\end{lemma}
\begin{proof}
By regularizing vertex degrees, e.g., by adding $\log\log n$ self-loops of weight $0$ at each vertex, we can assume that $\textup{vol}(B^\pm_{G\{V\setminus U^\pm\}}(s,r))\ge|B^\pm_{G\{V\setminus U^\pm\}}(s,r)|\log\log n$ for each ball computed on step~(\ref{item:2a}). Since we terminate the other algorithm immediately upon finding the smaller ball, this step takes $O(\textup{vol}(B^\pm_{G\{V\setminus U^\pm\}}(s,r+\epsilon\Delta/\log m)))=O(\textup{vol}(B^\pm_{G\{V\setminus U^\pm\}}(s,r)))$ time and adds $\textup{vol}(B^\pm_{G\{V\setminus U^\pm\}}(s,r))$ volume to $U^\pm$. Each of $U^+$ and $U^-$ supports $O(m+n\log\log n)$ volume after regularization, which also bounds the running time.
\end{proof}

\subsection{Heavy case}\label{sec:heavy-case}

In the heavy case, we have found a vertex $s\in V$ where the volume of the smaller ball $B^\pm_{G\{V\setminus U^\pm\}}(s,r)$ exceeds $m$. The ball in the other direction, if run to completion, also has volume more than $m$. Both balls have radius at most $\Delta$ by \Cref{lem:find-ball}, and they are on an induced subgraph of $G$. Increasing the radius to $\Delta$ and operating on $G$ instead, we conclude that $\textup{vol}(B^+_G(s,\Delta))>m$ and $\textup{vol}(B^-_G(s,\Delta))>m$.

The algorithm runs \Cref{lem:find-ball} on $G$ in the $+$ direction, and separately in the $-$ direction, both with initial radius $\Delta$ and additional limit $\Delta$. Let $r^+,r^-\in[\Delta,2\Delta]$ be the computed radii. The algorithm considers the three sets
\begin{align*}
X&=B^+_G(s,r^++\epsilon\Delta/\log m)\cap B^-_G(s,r^-+\epsilon\Delta/\log m),
\\Y&=B^+_G(s,r^++\epsilon\Delta/\log m)\setminus B^-_G(s,r^-), \\Z&=V\setminus B^+_G(s,r^+).
\end{align*}

\begin{lemma}\label{lem:size-heavy}
$X$ has weak diameter at most $6\Delta$. Moreover, $\textup{vol}(Y)\le m$ and $\textup{vol}(Z)\le m$ and $\textup{vol}(X)+\textup{vol}(Y)+\textup{vol}(Z)\le 2m+O(\epsilon m)$.
\end{lemma}
\begin{proof}
For any two vertices $u,v\in X$, there is a $u$-to-$s$ path in $G$ of weight at most $r^-+\epsilon\Delta/\log m\le3\Delta$, and an $s$-to-$v$ path of weight at most $r^++\epsilon\Delta/\log m\le3\Delta$, for a total of $6\Delta$. It follows that $X$ has weak diameter at most $6\Delta$.

To bound the volumes of $Y$ and $Z$, observe that $B^\pm_G(s,r^\pm)$ contains $B^\pm_G(s,\Delta)$ which has volume greater than $m$. Since $Y$ and $Z$ both exclude either $B^+_G(s,r^+)$ or $B^-_G(s,r^-)$, and since the total volume is $2m$, the bounds follow.

Finally, we show that $\textup{vol}(X)+\textup{vol}(Y)+\textup{vol}(Z)\le 2m+O(\epsilon m)$. It suffices to show that the vertices appearing in multiple sets have total volume $O(\epsilon m)$. Observe that $X\cup Y=B^+_G(s,r^++\epsilon\Delta/\log m)$, so any vertex in both $X\cup Y$ and $Z$ is in $B^+_G(s,r^++\epsilon\Delta/\log m)\setminus B^+_G(s,r^+)$, which has volume $O(\epsilon m)$ by the guarantee of \Cref{lem:find-ball}. Similarly, any vertex in both $X$ and $Y$ is in $B^-_G(s,r^-+\epsilon\Delta/\log m)\setminus B^-_G(s,r^-)$, which also has volume $O(\epsilon m)$.
\end{proof}

\begin{lemma}\label{lem:coloring-heavy}
Consider any path of weight $w$. The vertices of this path can be colored red, blue, and green such that
 \begin{enumerate}
 \item Each red vertex is in $X$, each blue vertex is in $Y$, and each green vertex is in $Z$.
 \item At most $8\lceil\frac w{\epsilon\Delta/\log m}\rceil$ edges on the path have differently colored endpoints.
 \end{enumerate}
\end{lemma}
\begin{proof}
We first color vertices \emph{yellow} and green such that each yellow vertex is in $X\cup Y$, and each green vertex is in $Z$. Since $X\cup Y=B^+_G(s,r^++\epsilon\Delta/\log m)$ and $Z=V\setminus B^+_G(s,r^+)$, we can proceed identically to \Cref{lem:coloring-light}, using the fact that any path segment from inside $B^+_G(s,r^+)$ to outside $B^+_G(s,r^++\epsilon\Delta/\log m)$ has weight exceeding $\epsilon\Delta/\log m$. We obtain a yellow-green coloring of the path with at most $2\lceil\frac w{\epsilon\Delta/\log m}\rceil$ edges having differently colored endpoints. In particular, there are at most $2\lceil\frac w{\epsilon\Delta/\log m}\rceil$ consecutive yellow segments.

We now re-color each consecutive yellow segment into red and blue. Again, since all vertices in $X$ are in $B^-_G(s,r^-+\epsilon\Delta/\log m)$ and all vertices in $Y$ are in $V\setminus B^-_G(s,r^-)$, we can apply \Cref{lem:coloring-light} again, obtaining at most $2\lceil\frac{w'}{\epsilon\Delta/\log m}\rceil\le2(\frac{w'}{\epsilon\Delta/\log m})+2$ red-blue edges per segment of weight $w'$. Since the weights $w'$ sum to at most $w$, and since there are at most $2\lceil\frac w{\epsilon\Delta/\log m}\rceil$ consecutive yellow segments, the total number of red-blue edges is at most $2(\frac w{\epsilon\Delta/\log m})+2\cdot2\lceil\frac w{\epsilon\Delta/\log m}\rceil$. Together with the green vertices, the overall number of differently colored edges is at most $8\lceil\frac w{\epsilon\Delta/\log m}\rceil$.
\end{proof}

\subsection{Combining both cases}

Finally, we conclude the proof of \Cref{thm:padded-decomposition}. In the light case, we have two sets $X,Y$ which we set as $X_1,X_2$, so $k=2$. In the heavy case, we have three sets $X,Y,Z$ which we set as $X_1,X_2,X_3$, so $k=3$. Both algorithms take $O(m+n\log\log n)$ time and fulfill property~(\ref{item:2}). For property~(\ref{item:1}), we set $\Delta=d/12$ so that $X$ has weak diameter at most $d/2$ in the heavy case by \Cref{lem:size-heavy}. All other sets have volume at most $1.6m$ by \Cref{lem:size-light,lem:size-heavy}, fulfilling property~(\ref{item:1}). Also, any path of weight at most $d=12\Delta$ has at most $8\cdot\lceil12\epsilon^{-1}\log m\rceil$ differently colored edges from \Cref{lem:coloring-light,lem:coloring-heavy}, which is at most $100\epsilon^{-1}\log m$ when $m$ is large enough, fulfilling property~(\ref{item:3}).

\bibliographystyle{alpha}
\bibliography{ref}

\end{document}